\documentclass[final,3p,times,twocolumn]{elsarticle}
\usepackage{epsfig}
\usepackage{amsmath}
\usepackage{color}
\usepackage{latexsym}
\usepackage{amssymb}
\usepackage{dsfont}
\usepackage{multirow}

\newcommand{\bea}{\begin{eqnarray}}
\newcommand{\eea}{\end{eqnarray}}
\newcommand{\be}{\begin{equation}}
\newcommand{\ee}{\end{equation}}

\newcommand{\ud}{\mathrm{d}}

\newlength\savedwidth

\newcommand{\uvec}[1]{\boldsymbol{#1}}

\begin{document}
\begin{frontmatter}

\author[orsay]{C.~Lorc\'e}
\author[pavia]{B.~Pasquini}

\address[orsay]{IPNO, Universit\'e Paris-Sud, CNRS/IN2P3, 
91406 Orsay, France\\
and LPT, Universit\'e Paris-Sud, CNRS, 91406 Orsay, France}
\title{The Pretzelosity TMD and Quark Orbital Angular Momentum}

\address[pavia]{Dipartimento di Fisica,
Universit\`a degli Studi di Pavia, Pavia, Italy, \\and Istituto Nazionale di Fisica Nucleare,
Sezione di Pavia, Pavia, Italy}

\begin{abstract}
We study the connection between the quark orbital angular momentum and the pretzelosity transverse-momentum dependent parton distribution function. We discuss the origin of this relation in quark models, identifying as key ingredient for its validity the assumption of spherical symmetry for the nucleon in its rest frame. Finally we show that the individual quark contributions to the orbital angular momentum obtained from this relation can not be interpreted as the intrinsic contributions, but include the contribution from the transverse centre of momentum which cancels out only in the total orbital angular momentum.
\end{abstract}

\end{frontmatter}

\section{Introduction}

One of the novel information contained in the transverse-momentum dependent parton distributions (TMDs) is the orbital motion of the partons inside the nucleon. Most of these TMDs would simply vanish in absence of quark orbital angular momentum (OAM). However, there exists no direct quantitative connection between the OAM and an observable related to TMDs. Any relation in this direction is bound to be model dependent.

Recently, it has been suggested, on the basis of some quark-model calculations, that the TMD $h_{1T}^\perp$ (also called \emph{pretzelosity} TMD due to the typical shapes it produces in the proton rest frame ~\cite{arXiv:0708.2297}) may be related to the quark OAM as follows~\cite{She:2009jq,Avakian:2010br,Efremov:2010cy}
\be\label{pretzelosity}
\mathcal L_z=-\int\ud x\,\ud^2\uvec k\,\frac{\uvec k^2}{2M^2}\,h_{1T}^{\perp}(x,\uvec k^2).
\ee
As emphasized in Ref.~\cite{Avakian:2010br}, the identification in Eq.~\eqref{pretzelosity} is valid at the amplitude level, but not at the operator level. Note also that the lhs is chiral even and charge even, while the rhs is chiral odd and charge odd. This means that Eq.~\eqref{pretzelosity} can only hold \emph{numerically} because of some simplifying assumptions in quark models. In this paper we review this relation in the context of quark models, elucidating its physical origin and the underlying model assumptions for its validity. 
\newline
The plan of the paper is as follows. In Section~\ref{sec:1} we introduce the representations for the quark OAM and the pretzelosity TMD in terms of overlap of light-cone wave functions (LCWFs), showing that they are in general different. As discussed in Section~\ref{sec:2}, the key ingredient to identify the two representations is to assume spherical symmetry for the nucleon in its rest frame. Within this assumption, we further show in Section~\ref{sec:3} that for the individual quark contributions the OAM calculated from the pretzelosity can not be identified with the intrinsic contribution. The difference is due to the contribution coming from the transverse centre of momentum which cancels out only in the total OAM. We conclude with a section summarizing our results.

\section{Overlap representation}
\label{sec:1}
In this section we discuss the validity of Eq.~\eqref{pretzelosity} comparing the overlap representations of the quark OAM and the pretzelosity TMD in terms of light-cone wave functions (LCWFs). 

In the nucleon Fock space, the $N$-parton state is described by the LCWF $\Psi^\Lambda_{N\beta}(r)$, with $\Lambda$ the nucleon helicity, the index $\beta$ labeling the quark light-cone helicities $\lambda_i$, flavours $q_i$, and colours. The LCWFs depend on $r=\{r_1,\dots,r_n\}$ which refers collectively to the momentum coordinates of the partons relative to the nucleon momentum, {\em i.e.} $r_i=(x_i,\uvec k_i)$ with $x_i$ the longitudinal momentum fraction and $\uvec k_i$ the transverse momentum. An important property of the LCWFs is that they are eigenstates of the total OAM (obtained from the sum over the $N$ parton contributions)
\be\label{OAMop}
-i\sum_{n=1}^N\left(\uvec k_n\times\uvec\nabla_{\uvec k_n}\right)_z\Psi^\Lambda_{N\beta}(r)=l_z\,\Psi^\Lambda_{N\beta}(r)
\ee
with eigenvalue $l_z=(\Lambda-\sum_n\lambda_n)/2$. As a consequence, the total OAM can  be simply expressed as
\begin{subequations}\label{rel}
\be
\mathcal L_z=\sum_{N,\,l_z}l_z\,\rho_{Nl_z},
\ee
where
\be
\label{eq:overlap}
\rho_{Nl_z}\equiv\sum_{\beta'}\delta_{l_zl'_z}\int\left[\ud x\right]_N\left[\ud^2\uvec k\right]_N\left|\Psi^+_{N\beta'}(r)\right|^2
\ee
\end{subequations}
is the probability to find the nucleon with light-cone helicity $\Lambda=+$ in an $N$-parton state with eigenvalue $l_z$ of the total OAM. In Eq.~\eqref{eq:overlap}, the integration measures are given by
\begin{align}
\left[\ud x\right]_N&=\left[\prod_{i=1}^N\ud x_i\right]\delta\left(1-\sum_{i=1}^N x_i\right),\\
\left[\ud^2\uvec k\right]_N&=\left[\prod_{i=1}^N\frac{\ud^2\uvec k_i}{2(2\pi)^3}\right]2(2\pi)^3\,\delta^{(2)}\left(\sum_{i=1}^N\uvec k_i\right).
\end{align}

The LCWF overlap representation of the rhs of Eq.~\eqref{pretzelosity} has been derived for the three-quark contribution in Refs.~\cite{Pasquini:2008ax,Lorce:2011dv,Lorce:2011zt} and can be generalized to the $N$-parton LCWF as
\begin{subequations}
\label{lzTMD}
\be
-\int\ud x\,\ud^2\uvec k\,\frac{\uvec k^2}{2M^2}\,h_{1T}^{\perp}(x,\uvec k^2)=\sum_{N,\,\beta}\sum_{n=1}^N\mathcal A^{N\beta}_n
\ee
with
\begin{align}
\label{ORfin}
\mathcal A^{N\beta}_n&=-\frac{1}{2}\sum_{\lambda'_n}\left(\sigma_L\right)_{\lambda'_n\lambda_n}\nonumber\\
&\times \int\left[\ud x\right]_N\left[\ud^2\uvec k\right]_N\,\hat k_{nR}^2\,\Psi^{*+}_{N\beta'}(r)\Psi^-_{N\beta}(r)
\end{align}
\end{subequations}
representing the contribution of the $n$th quark in the $N$-parton state. Here, $\beta'$ is the same as $\beta$ except that $\lambda_n$ is replaced by $\lambda'_n$. We used also the notations $\sigma_{R,L}=\sigma_x\pm i\sigma_y$ with $\sigma_i$ the Pauli matrices and $\hat k_{nR,L}=k_{nR,L}/|\uvec k_n|$ with $k_{nR,L}=k^x_n\pm ik^y_n$.

As one can see from the overlap representations~\eqref{rel} and \eqref{lzTMD}, the lhs of Eq.~\eqref{pretzelosity} involves no helicity flip and therefore no change of the total OAM ($|\Delta l_z|=0$), while the rhs involves two helicity flips in opposite directions (one at the quark level and one at the nucleon level) leading to a change by two units of total OAM ($|\Delta l_z|=2$). This clearly indicates that one should not expect Eq.~\eqref{pretzelosity} to hold in general. We will show however that non-mutually interacting quark models with spherical symmetry in the nucleon rest frame necessarily satisfy Eq.~\eqref{pretzelosity}.

\section{Spherically symmetric quark models}
\label{sec:2}

As argued in Ref.~\cite{Lorce:2011zt}, the models of Refs.~\cite{She:2009jq,Avakian:2010br,Efremov:2010cy} belong to the class of $SU(6)$ symmetric quark models where the three-quark (3Q) LCWFs can generically be written as
\be\label{LCWFs}
\Psi^\Lambda_{3\beta}(r)=\phi(r)\sum_{\sigma_1,\,\sigma_2,\,\sigma_3}\Phi^{\Lambda,q_1q_2q_3}_{\sigma_1\sigma_2\sigma_3}\prod_{n=1}^3D^{(1/2)*}_{\sigma_n\lambda_n}(r),
\ee
with $\phi(r)$ a symmetric momentum wave function normalized as $\int\left[\ud x\right]_3\left[\ud^2\uvec k\right]_3\left|\phi(r)\right|^2=1$, $\Phi^{\Lambda,q_1q_2q_3}_{\sigma_1\sigma_2\sigma_3}$ the $SU(6)$ spin-flavour wave function satisfying $\Lambda=\sum_n\sigma_n$, and $D^{(1/2)*}_{\sigma_n\lambda_n}(r)$ an $SU(2)$ Wigner rotation matrix relating the quark light-cone helicity $\lambda_n$ to the quark canonical spin $\sigma_n$ given by
\be\label{rot}
D^{(1/2)*}_{\sigma_n\lambda_n}(r)=\begin{pmatrix}\cos\tfrac{\theta(r)}{2}&-\hat k_{nR}\,\sin\tfrac{\theta(r)}{2}\\\hat k_{nL}\,\sin\tfrac{\theta(r)}{2}&\cos\tfrac{\theta(r)}{2}\end{pmatrix}_{\sigma_n\lambda_n}.
\ee
The angle $\theta(r)$ between the light-cone and canonical polarizations is usually a complicated function of the quark momentum $k$ and is specific to each model. The only general property is that $\theta\rightarrow 0$ as $\uvec k\rightarrow 0$. Note also that the general relation between light-cone helicity and canonical spin is usually quite complicated, as the dynamics is involved (see for example Ref.~\cite{Miller:2009fc}). Only in the case where the target is described in terms of quarks without mutual interactions, the LCWF can be cast in the form of Eq.~(\ref{LCWFs}).

Since the functions $\phi(r)$ and $\theta(r)$ depend on the quark transverse momenta only through $\uvec k^2_n$, one has
\begin{align}
\label{spherical}
&-i\left(\uvec k_n\times\uvec\nabla_{\uvec k_n}\right)_z
\left\{\begin{matrix}\phi(r)\\\theta(r)\end{matrix}\right\}\nonumber\\
&=\left(k_{nR}\,\frac{\partial}{\partial k_{nR}}-k_{nL}\,\frac{\partial}{\partial k_{nL}}\right)\left\{\begin{matrix}\phi(r)\\\theta(r)\end{matrix}\right\}=0.
\end{align}

Using the 3Q LCWF in Eq.~\eqref{LCWFs}, we find that the total quark OAM can be expressed as
\be\label{Lz}
\mathcal L_z=\int\left[\ud x\right]_3\left[\ud^2\uvec k\right]_3\left|\phi(r)\right|^2\,\sin^2\tfrac{\theta(r)}{2}
\ee
which implies that $0\leq\mathcal L_z<1$. Using now the results of Ref.~\cite{Lorce:2011zt} we find that the isosinglet axial charge $\Delta\Sigma=\sum_q\Delta q$ is given by
\be\label{AxialOR}
\Delta\Sigma=\int\left[\ud x\right]_3\left[\ud^2\uvec k\right]_3\left|\phi(r)\right|^2\left(1-2\,\sin^2\tfrac{\theta(r)}{2}\right).
\ee
It is then straightforward to check the conservation of the total angular momentum
\be
J_z=\tfrac{1}{2}\,\Delta\Sigma+\mathcal L_z=\tfrac{1}{2}.
\ee
The way the total angular momentum $J_z$ is shared between the quark OAM and spin contributions is driven only by the rotation \eqref{rot} which is a typical relativistic effect. In the nonrelativistic limit, there is no distinction between light-cone helicity $\lambda$ and canonical spin $\sigma$, and the spin rotation matrix \eqref{rot} reduces to the identity, with $\theta(r)\to 0$. One then recovers the well-known result that in a nonrelativistic picture, the nucleon spin arises only from the quark spins $J_z=\tfrac{1}{2}\,\Delta\Sigma^\text{NR}$ since $\mathcal L^\text{NR}_z=0$. Using again the results of Ref.~\cite{Lorce:2011zt}, we find for the rhs of Eq.~\eqref{pretzelosity}, 
\begin{align}
&-\int\ud x\,\ud^2\uvec k\,\frac{\uvec k^2}{2M^2}\,h_{1T}^{\perp}(x,\uvec k^2)
\nonumber\\
&=\int\left[\ud x\right]_3\left[\ud^2\uvec k\right]_3\left|\phi(r)\right|^2\sin^2\tfrac{\theta(r)}{2},
\end{align}
which agrees with the expression for $\mathcal L_z$ in Eq.~\eqref{Lz}.

As long as the LCWF keeps a structure similar to Eq.~\eqref{LCWFs}, one can proceed in the same way for any $N$-quark state and reach the same conclusion. We also note that the $SU(6)$ symmetry is not necessary as long as spherical symmetry in the nucleon rest frame is assumed.

\section{Flavour separation}
\label{sec:3}

Since the transverse-position operator $\hat{\uvec r}_n$ is represented in transverse momentum space by $i\uvec\nabla_{\uvec k_n}$, we may interpret $-i\left(\uvec k_n\times\uvec\nabla_{\uvec k_n}\right)_z$ as the operator giving the OAM contribution due to the $n$th quark. The total OAM can then be decomposed as
\be
\mathcal L_z=\sum_{N,\,\beta}\sum_{n=1}^N\mathcal L^{N\beta}_{nz},
\ee
where
\be\label{ORext}
\mathcal L^{N\beta}_{nz}=-i\int\left[\ud x\right]_N\left[\ud^2\uvec k\right]_N\,\Psi^{*+}_{N\beta}(r)\left(\uvec k_n\times\uvec\nabla_{\uvec k_n}\right)_z\Psi^+_{N\beta}(r).
\ee
represents the contribution of the $n$th quark in the $N$-parton state characterized by $\beta$. The contribution due to quarks of flavour $q$ is then given by
\be
\mathcal L^q_z=\sum_{N,\,\beta}\sum_{n=1}^N\delta_{qq_n}\,\mathcal L^{N\beta}_{nz}.
\ee
Note that this is the expression that was used in the model calculations of Refs.~\cite{She:2009jq,Avakian:2010br,Efremov:2010cy}.

Similarly, we have
\be
-\int\ud x\,\ud^2\uvec k\,\frac{\uvec k^2}{2M^2}\,h_{1T}^{\perp q}(x,\uvec k^2)=\sum_{N,\,\beta}\sum_{n=1}^N\delta_{qq_n}\,\mathcal A^{N\beta}_n
\ee 
with $\mathcal A^{N\beta}_n$ given by Eq.~\eqref{ORfin}. Using the generic 3Q LCWF in Eq.~\eqref{LCWFs}, we obtain the flavour-dependent version of Eq.~\eqref{pretzelosity}
\be
\mathcal L^q_z=-\int\ud x\,\ud^2\uvec k\,\frac{\uvec k^2}{2M^2}\,h_{1T}^{\perp q}(x,\uvec k^2).
\ee
We have therefore reproduced the result of Refs.~\cite{She:2009jq,Avakian:2010br,Efremov:2010cy} and extended its validity to the whole class of models where the 3Q LCWF has the generic structure \eqref{LCWFs}.

However, it is important to note that $\mathcal L^q_z$ does not represent the \emph{intrinsic} OAM contribution due to quarks of flavour $q$. While $\uvec k_n$ represents the intrinsic transverse momentum of the $n$th quark (we consider a target with vanishing transverse momentum, $\uvec P=\sum_n\uvec k_n=\uvec 0$), $i\uvec\nabla_{\uvec k_n}$ does not represent the intrinsic transverse-position operator of the $n$th quark in momentum space. In order to define an operator for the intrinsic transverse position, one has to specify a privileged point which is identified with the centre of the target. In a non-relativistic description, the centre of the target is identified with the centre of mass of the system. In a relativistic description, the transverse centre of the target $\uvec R$ is identified with the transverse centre of momentum $i\sum_jx_j\,\uvec\nabla_{\uvec k_j}$ in the light-front form \cite{Burkardt:2000za,Burkardt:2002hr,Burkardt:2005td}. The operator giving the intrinsic OAM contribution due to the $n$th quark is therefore $-i\sum_j(\delta_{nj}-x_j)\left(\uvec k_n\times\uvec\nabla_{\uvec k_j}\right)_z$. The intrinsic contribution of the $n$th quark in the $N$-parton state characterized by $\beta$ then reads
\begin{align}\label{ORint}
\ell^{N\beta}_{nz}&=-i\sum_{j=1}^N(\delta_{nj}-x_j)\nonumber\\
&\times\int\left[\ud x\right]_N\left[\ud^2\uvec k\right]_N\,\Psi^{*+}_{N\beta}(r)\left(\uvec k_n\times\uvec\nabla_{\uvec k_j}\right)_z\Psi^+_{N\beta}(r).
\end{align}
Note that the expression for $\ell^q_z=\sum_{N,\,\beta}\sum_n\delta_{qq_n}\,\ell^{N\beta}_{nz}$ coincides with the OAM computed directly from the quark phase-space distribution \cite{Lorce:2011kd,Lorce':2011ni}
\be
\ell^q_z=\int\ud x\,\ud^2\uvec k\,\ud^2\uvec r\left(\uvec r\times\uvec k\right)_z\,\rho^{[\gamma^+]q}_{++}(\uvec r,\uvec k,x),
\ee
where $\rho^{[\gamma^+]q}_{++}(\uvec r,\uvec k,x)$ is the Wigner distribution for unpolarized quarks with flavour $q$ in a longitudinally polarized nucleon. This supports the interpretation of the Wigner functions defined in Ref.~\cite{Lorce:2011kd} as intrinsic phase-space distributions of quarks inside the target. Equation \eqref{ORint} also corresponds to the LCWF overlap representation of the Jaffe-Manohar OAM~\cite{Lorce':2011ni}.

In general we have $\ell^q_z\neq\mathcal L^q_z$ while $\ell_z=\mathcal L_z$, as it was observed in Ref.~\cite{Lorce:2011kd} for the light-cone version of the chiral quark-soliton model and a light-cone constituent quark model. This can be understood with simple classical arguments. If the coordinates of the $n$th parton with respect to some origin $O$ are $\uvec r_n$, the coordinates of the same parton with respect to another origin $O'$ are $\uvec r'_n=\uvec r_n-\uvec d$, where $\uvec d$ is the vector connecting the two origins in the transverse plane. We then have
\be
\sum_n \uvec r'_n\times\uvec k_n=\sum_n \uvec r_n\times\uvec k_n-\uvec d\times\sum_n\uvec k_n=\sum_n \uvec r_n\times\uvec k_n,
\ee
since $\sum_n\uvec k_n=\uvec P=\uvec 0$. In other words, the fact that the nucleon has no transverse momentum removes the dependence of the total OAM on the choice of the privileged point. One can also directly see that once summed over $n$, Eqs.~\eqref{ORext} and \eqref{ORint} become identical.

\section{Conclusions}

We showed that, for a large class of quark models based on spherical symmetry, the orbital angular momentum can be accessed \emph{via} the pretzelosity transverse-momentum dependent parton distribution. The individual quark contributions to the orbital angular momentum obtained in this way do however not correspond to the intrisic quark orbital angular momentum. On the other hand, the intrinsic contributions can be obtained from the Wigner distributions as recently shown in Ref.~\cite{Lorce':2011ni}. The two calculations agree for the total OAM, since in the sum over the individual quark contributions the spurious terms due to the transverse centre of momentum cancel out.

\section*{Acknowledgements}
We thank P. Schweitzer for stimulating comments. C. L. is thankful to INFN and the Department of Physics of the University of Pavia for their hospitality. This work was supported in part by the European Community Joint Research Activity ``Study of Strongly Interacting Matter'' (acronym HadronPhysics3, Grant Agreement n. 283286) under the Seventh Framework Programme of the European Community, and by the Italian MIUR through the PRIN 2008EKLACK ``Structure of the nucleon: transverse momentum, transverse spin and orbital angular momentum'' and by the P2I (``Physique des deux Infinis'') project.

\end{document}